\documentstyle[12pt]{article}
\begin{document}
\begin{titlepage}

\pagestyle{empty}

\begin{flushright}
{\footnotesize UFRN-DFTE\\

August 2001}
\end{flushright}

\vskip 1.0cm

\begin{center}
{\Large \bf Note on Solving for the Dynamics of the Universe}

\vskip 1cm
Jos\'{e} Ademir Sales Lima  
\end{center}
\vskip 0.5cm

\begin{quote}
\begin{center}
Universidade Federal do Rio Grande do Norte\\
 Departamento de F\'{\i}sica, C. P. 1641\\
     59072 - 970, Natal, RN, Brazil
\end{center} 
\end{quote}

\vskip 1.0cm

\centerline{\bf ABSTRACT}
\bigskip

\noindent In a recent article, Faraoni proposed an alternative procedure to solve the Friedman-Lemaitre-Robertson-Walker (FLRW) cosmological equations. The basic result of that paper was obtained long ago through a different approach, which seems to be little known and deserves closer attention due to its pedagogical interest. The broad importance of this method is readily recognized by examining some additional cases not considered by Faraoni. Its instructive potential for introductory courses on cosmology has been positively verified by the author in many lectures during the last decade.
\end{titlepage}

\newpage
\pagestyle{plain}
\baselineskip 0.7cm
\newpage

Recently, Faraoni\cite{F99} rediscussed the cosmological solutions for a relativistic simple fluid in the framework of FLRW models. In this note we present a variant of the method proposed by Faraoni, which seems to be more general and quite interesting from a pedagogical viewpoint. Although little known, this approach was proposed long ago \cite{AL88}, and its pedagogical efficiency has also been verified many times in the last decade. In what follows, after a comparison with the results presented by Faraoni, we stress the generality of this procedure by discussing a class of solutions with a cosmological constant. For the sake of completness and further reference, as well as to consider some subtleties not addressed in Faraoni's work, we first summarize the basics of this problem.

The spacetime metric takes the following form ($c=1$)

\begin{equation}
\label{M}
ds^2=dt^2-a(t)^2\left(\frac{dr^2}{1-Kr^2} + r^2 d{\theta}^{2} + r^2 sin^{2}{\theta}d{\phi}^{2}\right)\,
\end{equation} 
where $a(t)$ is the scale factor, and $K=0,\pm1$ is the curvature parameter. In such a background, the Einstein field equations  for 
a relativistic simple fluid can be written as [1-5]

\begin{equation}
\label{rho} 
8\pi G\rho  = 3\frac{{\dot a}^2}{a^2} +
3\frac{K}{a^2}\ , 
\end{equation}
\begin{equation}
\label{p}
8\pi Gp = -2\frac{{\ddot a}}{a} -\frac{{\dot
a}^2}{a^2}-\frac{K}{a^2}\ ,
\end{equation} 
where $\rho$ and $p$ are the energy density and pressure, respectively. In this system there are three unknown quantities, namely $a(t)$, $\rho(t)$ and $p(t)$ and only two independent equations. Thus, in order to solve it, an additional constraint is required. In the cosmological framework, it is usually assumed that the matter content obeys the barotropic $\gamma$-law equation of state

\begin{equation}
\label{ES} 
p=(\gamma -1)\rho\ , \ \ \ \ \ \ \ \ \ \gamma \in
[0,2]\quad. 
\end{equation}
This happens because such an expression provides a rather simple description 
for several cases of physical interest \cite{ZN96}: (i) dust or pressureless
matter  ($\gamma = 1$), (ii) radiation ($\gamma = 4/3$), (iii) vacuum
($\gamma = 0$), and (iv) Zeldovich's stiff matter ($\gamma = 2$). Naturally,
the above assumption does not imply that the $\gamma-parameter$ remains
constant in the course of the cosmological evolution. In the standard hot big
bang description, the Universe underwent a transition from a primordial
radiation phase $(\gamma= 4/3)$ to a matter-dominated epoch ($\gamma =1$)
during the late stages of its evolution. 

The cosmic dynamics is determined by combining the above set of equations. 
In principle, the corresponding dynamic behavior must be heavily dependent on
the choice of the two free parameters: (i) the curvature parameter K, and (ii)
the equation of state parameter $\gamma$. As one may check, the evolution of
the scale function is driven by the second order differential
equation\cite{F99,AL88}

\begin{equation}
\label{FRW} 
a{\ddot a} +\Delta {\dot a}^2 +\Delta K =0 \quad, 
\end{equation} 
where the parameter $\Delta$ ($c$ in Faraoni's notation) is a function of $\gamma$ 

\begin{equation}
\label{D}
\Delta = {\frac{3\gamma - 2}{2}} \quad.
\end{equation}

The first integral of (\ref{FRW}) is 
\begin{equation}
\label{I1} 
{\dot a}^2 = \left({\frac{a_o}{a}}\right)^{2\Delta} - K \quad, 
\end{equation} 
where the convenient integration constant, ${a_o}^{2\Delta}$, requires that
all curves  described by the general solution share the point $a=a_o$ at some
instant of the cosmological time. Furthermore, the derivative  ${\dot a}(a_o)=
\sqrt {1 - K}$ is the same for all values of $\gamma$, and indicates the
existence of an extremum for closed models. We see from (\ref{FRW}) and
(\ref{D}) that this extremum is a maximum if $\gamma > {2 \over 3}$, a minimum
if $\gamma < {2 \over 3}$, or a stationary point if $\gamma = {2 \over 3}$
\cite{C1}. In addition, equation  (\ref{rho}) tells us that the energy density
\begin{equation} \label{rhos} 
\rho = {3 \over
{a_o}^{2}}\left(\frac{a_o}{a}\right)^{3\gamma}\quad, 
\end{equation}  
is positive definite, as should be expected for a physical
fluid. Note also that for $K=0$ the first integral (\ref{I1}) has the
following solution

\begin{equation}
\label{flat1}
a(t)=a_o\left[1 + (\Delta + 1)(t-t_o)/a_o \right]^{1 \over {\Delta + 1}},
\end{equation}
or equivalently,
\begin{equation}
\label{flat2}
a(t)=a_o\left[1 + {3\gamma\over 2}({{t-t_o} \over a_o}) \right]^{{2 \over 3\gamma}},
\end{equation}
where $t_o$ is an arbitrary time scale. In particular, by choosing $t_o =
{2a_{o}/3\gamma}$  one recovers the restricted form of the flat solutions,
namely, $a(t)=a_o (t/t_o)^{2 \over 3\gamma}$. In addition, in the limit
$\gamma \rightarrow 0$, the above expression  reduces to $a(t) \sim 
e^{{H_o}t}$, which is the de Sitter flat solution ($H_o=a_o^{-1}$).

At this point we remark that the aim of Faraoni's paper is to obtain the
general solution  of the first integral (\ref{I1}), or equivalently of
(\ref{FRW}). This can be accomplished by using the conformal time $\eta$,
instead of the cosmological or physical time ($dt = a(\eta)d\eta$). In this
case, the metric (\ref{M}) and equation (\ref{FRW}) takes the forms below
\begin{equation} \label{M1} 
ds^2=a(\eta)^2\left(d{\eta^2}-\frac{dr^2}{1-Kr^2}
- r^2 d{\theta}^{2} - r^2 sin^{2}{\theta}d{\phi}^{2}\right)\, 
\end{equation}
\begin{equation} \label{FRW1}  
aa'' + (\Delta - 1) {a'}^2 + \Delta Ka^{2} = 0 \quad,  
\end{equation} 
where the prime denotes derivative with respect to conformal time.  Now,
instead of using Faraoni's transformation, $u={a'/a}$, which leads to
Ricatti's differential equation (see section 3 of \cite{F99}), we employ the
auxiliary scale factor \cite{AL88} 

\begin{eqnarray}
& & Z(\eta) = a^{\Delta},  \,\,\,\, if \,\,\,\,{\Delta \neq 0}\, \label{Z1}\\
\nonumber\\
& & Z(\eta) = \ln {a}, \,\,\,\,\,\, if \,\,\,{\Delta = 0} \quad, \label{Z2}
\end{eqnarray}

to obtain, respectively,  

\begin{eqnarray}
& &{Z'' + K{\Delta}^{2}Z = 0}, \,\,\,\, if \, \,\,\, {\Delta \neq 0}\, \label{Z3}\\
\nonumber \\
& &{Z'' = 0},  \,\,\,\,\,\,\,\, if \,\,\,\, {\Delta = 0} \quad. \label{Z4}
\end{eqnarray} 

Equation (\ref{Z3}) is identical to equation (3.7) in Faraoni's paper. Note
that it reduces to (\ref{Z4}) in the limiting case $\Delta = 0$. Therefore, it
can be considered the general equation of motion for any value of $\Delta$. As
remarked before, it depends strongly on the pair of parameters ($\gamma, K$).
The physical meaning of (\ref{Z3}) is apparent: It describes the classical
motion of a particle subject to a linear force. This force is of restoring or
repulsive type depending on the sign of the curvature parameter. 

Closed models (K=1) are, for any value of $\Delta \neq 0$, analogous to 
harmonic simple oscillators (HSO).  Therefore, the cosmic dynamics in this
case is similar to a spring-mass system where the spring constant is
determined by the $\gamma$-parameter defining the equation of state obeyed by
the cosmic fluid (see Eq.(\ref{D})). It is worth noticing that for positive
values of $\Delta$, that is, $\gamma > 2/3$, this general and intermittent
oscillatory motion between both singularities (``big-bang" and ``big-crunch")
reinforces the connection with the idea of a pulsating Universe. From a
pedagogical viewpoint, these harmonic solutions for closed models are more
enlightening than the cycloidal parametric solutions usually presented in 
textbooks for particular values of $\gamma$ [3-5]. It is actually surprising
that the dynamics of closed universes might be reduced to that of a simple
harmonic oscillator, a key system for analytical and algebraic calculations in
many disparate branches of physics. It also leads naturally to the question
whether our Universe is the largest clock, that is, a global oscillator.
Indeed, although the current literature (including many textbooks) admits only
a unique incomplete cycle to describe the observed Universe, the existence of
pulsating solutions fascinate many philosophers and cosmologists since they
are associated with the old concept of an ethernal return.

It is also worth mentioning that if the Universe is spatially flat ($K=0$),
equation  (\ref{Z3})  implies that the system behaves like a free particle,
and the same happens if $\Delta=0$. In the later case, this effective free
particle behavior holds regardless of the curvature parameter.  For hyperbolic
spacetimes ($K=-1$), the system behaves like an ``anti-oscillator", that is, a
particle subject to a repulsive force proportional to the distance.

Now, recalling the elementary mathematical identities: 
\begin{eqnarray}
& &\lim_{\alpha \rightarrow 0} \frac{sin{\alpha x}}{\alpha} = x\, \\
& &sin({ix})= isinhx \quad, 
\end{eqnarray}
the unified solution of (\ref{Z3}) can be written as 

\begin{equation}
Z=\frac{Z_o}{\sqrt{K}}sin\sqrt{K}\left[|\Delta|(\eta + \delta\right)]
\end{equation}
where $Z_o$ and $\delta$ are integration constants. Note also that using the
transformation  (\ref{Z1}), the first integral (\ref{I1}) may be translated
into the energy equation 
\begin{equation} 
{1 \over 2}{Z'}^{2} + {1 \over 2}K{\Delta}^{2}Z^{2} = {1 \over
2}{\Delta}^{2}{a_o}^{\Delta} 
\end{equation} 
which determines $Z_o = {a_o}^{\Delta}$. 

By choosing $\delta=0$, the general parametric  solution relating the scale
factor and the cosmological time, $dt = a(\eta)d\eta$, is given by 
\begin{eqnarray} 
& &a(\eta) = a_o\left(\frac{sin\sqrt{K}|\Delta|\eta}{\sqrt
K}\right)^{1 \over \Delta},  \label{S1}\\ & &t(\eta) = a_o{\int
\left(\frac{sin\sqrt{K}|\Delta|\eta}{\sqrt K}\right)^{1 \over \Delta}} +
\mbox{constant} \quad. \label{S2} 
\end{eqnarray} 
From identity (18) these solutions for elliptic and hyperbolic models read:

$K=1$
\begin{eqnarray}
& &a(\eta) = a_o\left(sin|\Delta|\eta \right)^{1 \over \Delta}\label{S5}\\ 
& &t(\eta) = a_o{\int \left(sin|\Delta|\eta \right)^{1 \over \Delta}} +
\mbox{constant} \label{S5a} 
\end{eqnarray}

$K = -1$
\begin{eqnarray}
& &a(\eta) = a_o\left(sinh|\Delta|\eta\right)^{1 \over \Delta}\label{S3}\\
& &t(\eta) = a_o{\int \left(sinh|\Delta|\eta \right)^{1 \over \Delta}} +
\mbox{constant} \label{S3a} 
\end{eqnarray}
where the integrals (\ref{S5a}) and (\ref{S3a}) for $t(\eta)$ may be 
represented in terms of hypergeometric  Gaussian functions \cite{AL88}. As
one may check using (17), the integral for $K=0$ is trivial, and the
parametric solutions may readily be inverted to give the scale factor as a
function of the cosmological time.

It is also interesting to show that the above method based on the
transforming equations (\ref{Z1}) and (\ref{Z2}) is also convenient when new
ingredients are considered such as the presence of a cosmological 
$\Lambda$-term \cite{L90}. In this case, the EFE equations can be written as 

\begin{equation}
\label{rho1} 
8\pi G\rho + \Lambda  = 3\frac{{\dot a}^2}{a^2} +
3\frac{K}{a^2}\ , 
\end{equation}
\begin{equation}
\label{p1}
8\pi Gp - \Lambda =-2\frac{{\ddot a}}{a} -\frac{{\dot
a}^2}{a^2}-\frac{K}{a^2}\ ,
\end{equation} 
where $\Lambda$ is the cosmological constant. Now, using (\ref{ES}), one 
may see that the scale factor satisfies the generalized FLRW equation 
\begin{equation}
\label{FRWM} 
a{\ddot a} +\Delta {\dot a}^2 +\Delta K - {1 \over 3}\Delta(\Delta + 1)\Lambda a^{2}=0 \quad. 
\end{equation} 

Following the same steps of the case with  no $\Lambda$, instead of
(\ref{FRW1}), we obtain in the conformal time 
\begin{equation} \label{FRW2} 
aa'' + (\Delta - 1) {a'}^2 + \Delta Ka^{2} - {1 \over 3}\Delta(\Delta + 1)\Lambda a^{4}= 0  \quad, 
\end{equation} 
and transforming to the auxiliary scale factor through (\ref{Z1}) results

\begin{equation}
\label{FRW3}
Z'' + K{\Delta}^ {2}Z - {1 \over 3}\Delta(\Delta + 1)\Lambda Z^{{\Delta + 2}
\over \Delta} = 0  \quad.  
\end{equation}
As expected, if $\Lambda = 0$,  equations (\ref{FRWM}), (\ref{FRW2}) and
(\ref{FRW3}) reduce to (\ref{FRW}), (\ref{FRW1}) and (\ref{Z3}), respectively.

Physically,  the equation of motion (\ref{FRW3}) means that closed universes
with cosmological constant evolve like anharmonic or non-linear oscillators
\cite{LLM}.  The anharmonic contribution to the HSO is proportional to the
cosmological $\Lambda$-term, and its power index depends uniquely on the
equation of state $\gamma$-parameter. Note that the method provides an exact
non-linear description. In other words, the anharmonic term does not appear
due to the retention of higher order terms in some particular perturbative
scheme.   

Finally, we remark that the unidimensional  equation of  motion (\ref{FRW}),
as well as its generalized version with a cosmological constant are endowed 
with an obvious mechanical analogy which can be described within the classical
Lagrangian formalism. It has been shown that a particle subject to a potential
$V(a) = \alpha{a^{-n}}- {1 \over 6}\Lambda a^{2}$, where $\alpha$ and $n$ are
constants, may satisfy the equation of motion (\ref{FRWM}). More precisely,
this happens if $n = 2\Delta$, and the energy is proportional to the
curvature, $2E = -mK$, where $m$ is the mass of the test particle
\cite{LMJ98}. Naturally, the same method of solution, shown here for the
relativistic case, may also be applied to this classical Lagrangian approach.
The author has explored both approaches in many introductory courses on
cosmology, and verified the nice pedagogical impact on the students.   

\section*{Acknowledgments}
This work was partially supported by the CNPq 
(Brazilian research agency).


\begin{thebibliography}{99}
\bibitem{C1} Actually, the case $\Delta = 0$ should be separately discussed
because from (\ref{FRW}) one has $\ddot a(t) = 0$. This means that the
solution, $a(t) = a_o +  b_ot$, is valid for all values of $K$. Taking the
first integral as given by (\ref{I1}), the particular closed expansionist
solution ($b_o \neq 0$) is lost.
\end{thebibliography}
\end{document}